\documentclass[12pt]{iopart}

\usepackage{iopams}  

\usepackage{graphicx}

\begin{document}

\title{All-optical mode conversion via spatially-multimode four-wave mixing}

\author{Onur Danaci$^1$, Christian Rios$^1$ and Ryan T. Glasser$^{1,*}$}

\address{$^1$Department of Physics and Engineering Physics, Tulane University, 6400 Freret Street, New Orleans, Louisiana 70118 USA}

\ead{$^*$rglasser@tulane.edu} 

\vspace{10pt}
\begin{indented}
\item[]Aug. 03, 2016
\end{indented}

\begin{abstract}
We experimentally demonstrate the conversion of a Gaussian beam to an approximate Bessel-Gauss mode by making use of a non-collinear four-wave mixing process in hot atomic vapor.  The presence of a strong, spatially non-Gaussian pump both converts the probe beam into a non-Gaussian mode, and generates a conjugate beam that is in a similar non-Gaussian mode. The resulting probe and conjugate modes are compared to the output of a Gaussian beam incident on an annular aperture that is then spatially filtered according to the phase-matching conditions imposed by the four-wave mixing process.  We find that the resulting experimental data agrees well with both numerical simulations, as well as analytical formulae describing the effects of annular apertures on Gaussian modes.  These results show that spatially-multimode gain platforms may be used as a new method of mode conversion.
\end{abstract}

\section{Introduction}
Light propagating in non-Gaussian spatial modes has gained significant interest in recent history, and has been shown to be a useful tool in a variety of classical, nonlinear, and quantum optics schemes \cite{durnin_diffractionfree_1987,gori_besselgauss_1987,sheppard1977use,cruzramirez_observation_2012,mclaren_entangled_2012,lekki_singlephoton_2004}.  For example, making use of modes that exhibit nonzero orbital angular momentum has resulted in dramatic increases in information transfer rates, as well as allowed for the demonstration of high-dimensional entanglement \cite{fickler,romero,mclaren_selfhealing_2014,mclaren_twophoton_2013,dada_experimental_2011}.  It has also been shown that entangled states, when projected into Bessel-Gauss modes, are more robust to propagation than those in typical Gaussian modes \cite{durnin_diffractionfree_1987,durnin_exact_1987}.  Additionally, both Bessel-Gauss and Airy modes exhibit self-healing and limited diffraction properties, which have straightforward applications in optical communications and imaging \cite{besselreview,herman1,Muys,mclaren_selfhealing_2014}.  In this manuscript, we present experimental results in which nondegenerate four-wave mixing is used to both convert an input Gaussian probe mode to a Bessel-Gauss beam, while simultaneously generating a spatially-separate Bessel-Gauss conjugate beam.  

Nondegenerate four-wave mixing (4WM) in warm atomic vapor has proven to be a diverse tool in quantum optics \cite{slusher_observation_1985,Slusher_87,kumar_1994,mccormick_strong_2007,boyer_generation_2008,mccormick_strong_2008,tunabledelay,pooser2011,albertoPRL_2008,lownoise_2009,clark_quantum_2014}.  In this third-order nonlinear process, a pair of photons from a strong pump beam are annihilated, and a pair of photons are created at a small angle relative to the pump (one in the ``probe" mode and one in the ``conjugate" mode).  The angles and frequencies of the resulting photons are determined by energy-conservation and phase-matching conditions.  Due to the strong correlation between the resultant probe and conjugate modes, a large degree of sub-shot-noise squeezing has been demonstrated with bright beams, which may be used, for example, to enhance the sensitivity in metrological measurements \cite{jjin_2011,wang_2015,hudelist_2014,assaf1,optica_2015,otterstrom_14,acs_2016}.  The process, when not seeded with a probe beam, has been shown to result in strongly entangled output beams.  As this process is inherently multi-spatial-mode, squeezed and entangled images have also been demonstrated \cite{boyer_entangled_2008}. Additionally, due to this multi-spatial-mode nature \cite{boyer_generation_2008}, the 4WM process has been shown to allow the simultaneous generation of multiple pairs of entangled beams, applicable to quantum communication protocols \cite{gupta2016}.   

In the aforementioned 4WM experiments, the pump beam is always, to the best of our knowledge, larger in  spatial extent than the input probe beam.  This is typically desirable, as it results in nearly uniform amplification of the input probe mode profile.  Here we investigate the behavior of such a 4WM system when both the input probe is larger spatially than the pump beam, and the pump mode profile is non-Gaussian.  We find that this setup allows for the conversion of the probe Gaussian mode into non-Gaussian, Bessel-Gauss-like modes when the pump has an annular, donut-like profile.  This is due to limiting the interaction region that the probe experiences, resulting in an effective gain-medium aperture. Contrary to other methods of generating these non-Gaussian modes, the present scheme amplifies the input beam, rather than severely attenuate the input power (in the case of a transmissive diffraction grating, an annular aperture and lens combination, or less so when using an axicon lens \cite{besselreview}), while allowing for higher-powers to be used (as opposed to using, for example, spatial-light modulators which are easily damaged at higher-powers).

\begin{figure}[b]
\centering\includegraphics[width=13cm]{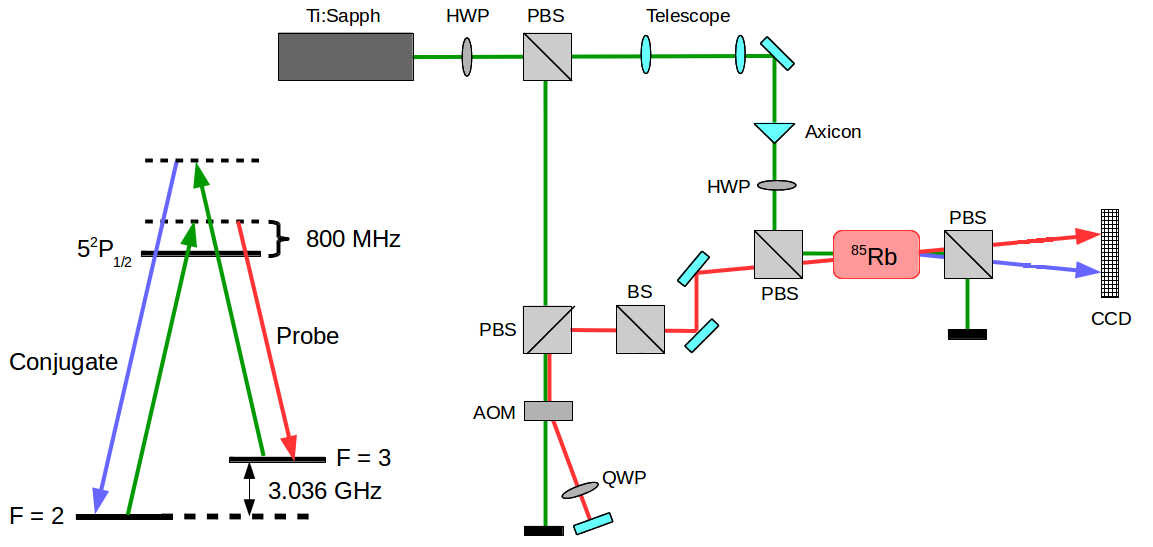}
\caption{Double-lambda level scheme (left) and experimental setup (right).  HWP: half-wave plate; PBS: polarization beam splitter; BS: non-polarizing beam splitter; AOM: acousto-optic modulator; QWP: quarter-wave plate.  After the four-wave mixing process the probe and conjugate are either directly incident onto a CCD (pixel pitch of 5.5 $\mu$m, or sent through a pair of lenses to collimate the beams and then investigate the modes'	 Fourier transform behavior.  Not shown is that an annular aperture and variable slit are placed at the location of the rubidium cell (and the pump beam is blocked), in order to compare the resulting mode profiles to those of the 4WM modes.}
\end{figure}

\section{Experimental setup}
The non-collinear 4WM scheme used here is shown in Fig. [1], and is similar to that described in \cite{mccormick_strong_2007}.  The setup consists of a strong (550 mW) pump beam that is detuned $\sim$ 800 MHz to the blue of the $^{85}$Rb D1 $F_{g} = 2$ to $F_{e} = 3$ transition.  The pump spatial mode in the 4WM interaction region is approximately donut-shaped, with a peak-to-peak diameter of $\sim$ 430 $\mu$m.  It is generated by placing an axicon lens in the pump beam path, and placing the nonlinear medium (1.7 cm long rubidium-85 vapor cell) before the axicon focal region \cite{axicon1,paulikas}.  The rubidium cell is heated to and held at a constant temperature of $\sim$ 115$^o$C.  A weak (50 $\mu$W) Gaussian spatial-mode probe beam detuned by 3.024 GHz to the red of the pump beam is injected at an angle of $\sim$ 1$^o$ with respect to the pump.  The full-width at half-maximum of the Gaussian probe in the interaction region is 1 mm.  The probe beam is generated by double-passing a high-frequency acousto-optic modulator and coupling the output into a single-mode fiber, allowing for precise control over its detuning relative to the pump beam.  The pump and probe beam are orthogonally polarized relative to one another (pump vertical and probe horizontal), combined on a polarizing beam splitter before the rubidium cell, and separated by another polarizing beam splitter after the interaction.                                                     

Due to conservation of energy and phase-matching conditions, the 4WM process generates a conjugate beam that is detuned to the blue side of the pump beam, and propagates at the same angle as the probe beam, on the other spatial side of the pump.  After the amplification of the probe, and generation of the conjugate beam, we place a CCD camera to examine the spatial-mode profiles of the resultant beams, with and without spatially Fourier transforming them. 

\begin{figure}[b]
\centering\includegraphics[width=13cm]{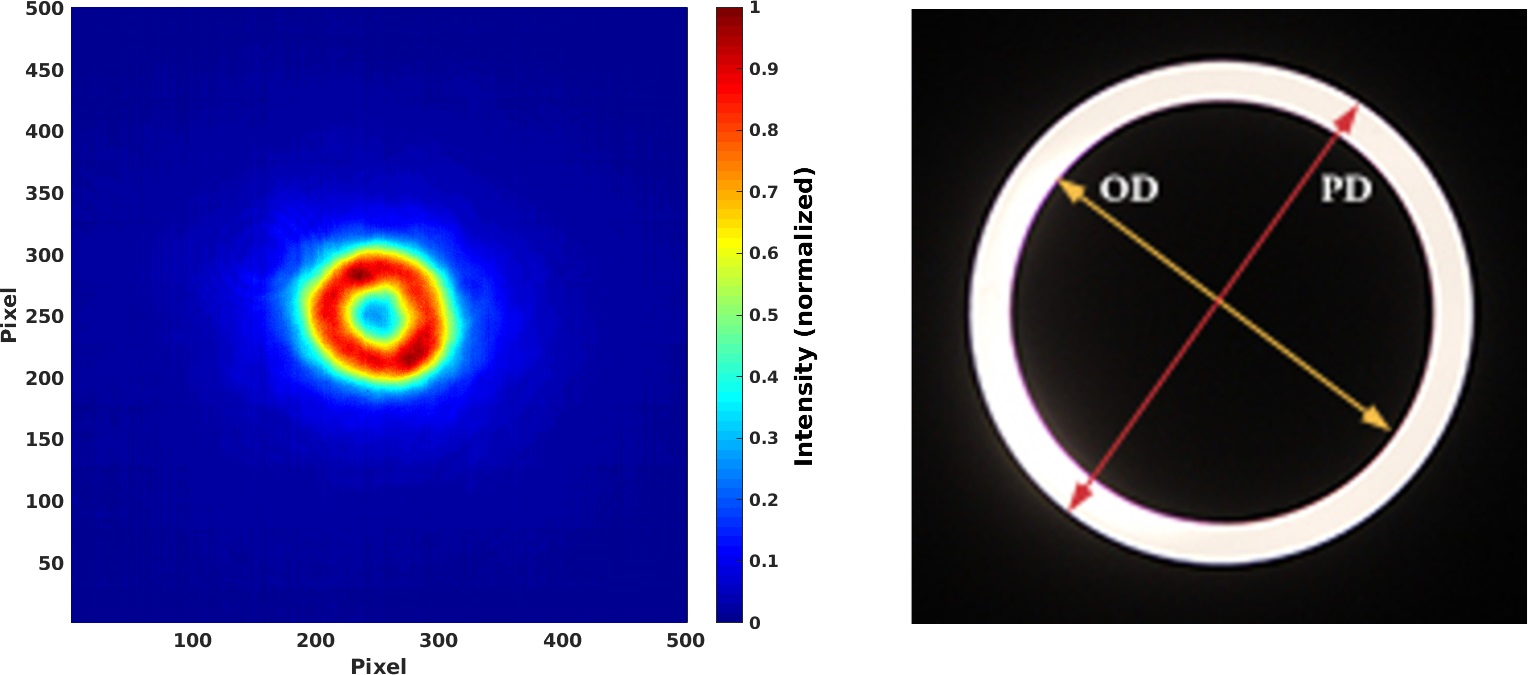}
\caption{Pump mode profile (left) in the interaction region and a schematic of the annular aperture (right) used to mimic the effect of the 4WM gain aperturing.}
\end{figure}

Contrary to all previous 4WM experiments, the pump beam here is smaller in spatial size than the injected probe beam, and has a non-Gaussian, donut-like spatial structure.  As a result, we find that the 4WM process acts as an effective aperture, limited by the finite spread of k-vectors supported by the phase-matching condition of the process, in combination with the effective interaction time mismatch in the vertical and horizontal directions due to the non-collinearity of the setup.  While a typical aperture serves to block some portion of the incident radiation, the 4WM process produces photons only along the directions within the phase-matching cone of angles, and where there is spatial overlap of the pump and probe beams.  Thus, in the high-gain regime (gain of $\sim$ 30), the input probe mode is effectively converted into the appropriate non-Gaussian mode as if it had passed through an annular aperture (when using an annular pump beam, as in the present experiment).  Additionally, we find that the generated conjugate beam is created in a similarly non-Gaussian mode.

In order to test the effective aperturing due to the 4WM process, we compare the output mode profiles (and their Fourier transforms) to those generated by passing the probe beam through a suitably-sized annular aperture.  Additionally, we place a tunable slit after the annular aperture to simulate the reduction in interaction time in the horizontal direction, as the pump has a null in intensity in its central region.  Thus, the top and bottom portion of the probe interact with the pump beam for a longer time than the horizontal components (as the latter pass through the null in the pump, as the beams are overlapped at a small angle in the horizontal direction).

\section{Results and discussion} 
The inherently multi-spatial mode nature of the 4WM process used here allows for the amplification of, for example, images that are imparted on the input probe beam.  However, difficulties in generating high-power non-Gaussian modes to be used as the pump beam in such experiments has resulted in nearly all experiments making use of Gaussian pump mode profiles. This results in the straightforward nearly-uniform amplification of input probe modes, and generation of conjugate modes that mimic (albeit mirrored) that of the probe profile.  A notable exception to this is an experimental demonstration that the 4WM process conserves orbital angular momentum, as shown in \cite{delocalized}.

\begin{figure}[b]
\centering\includegraphics[width=13cm]{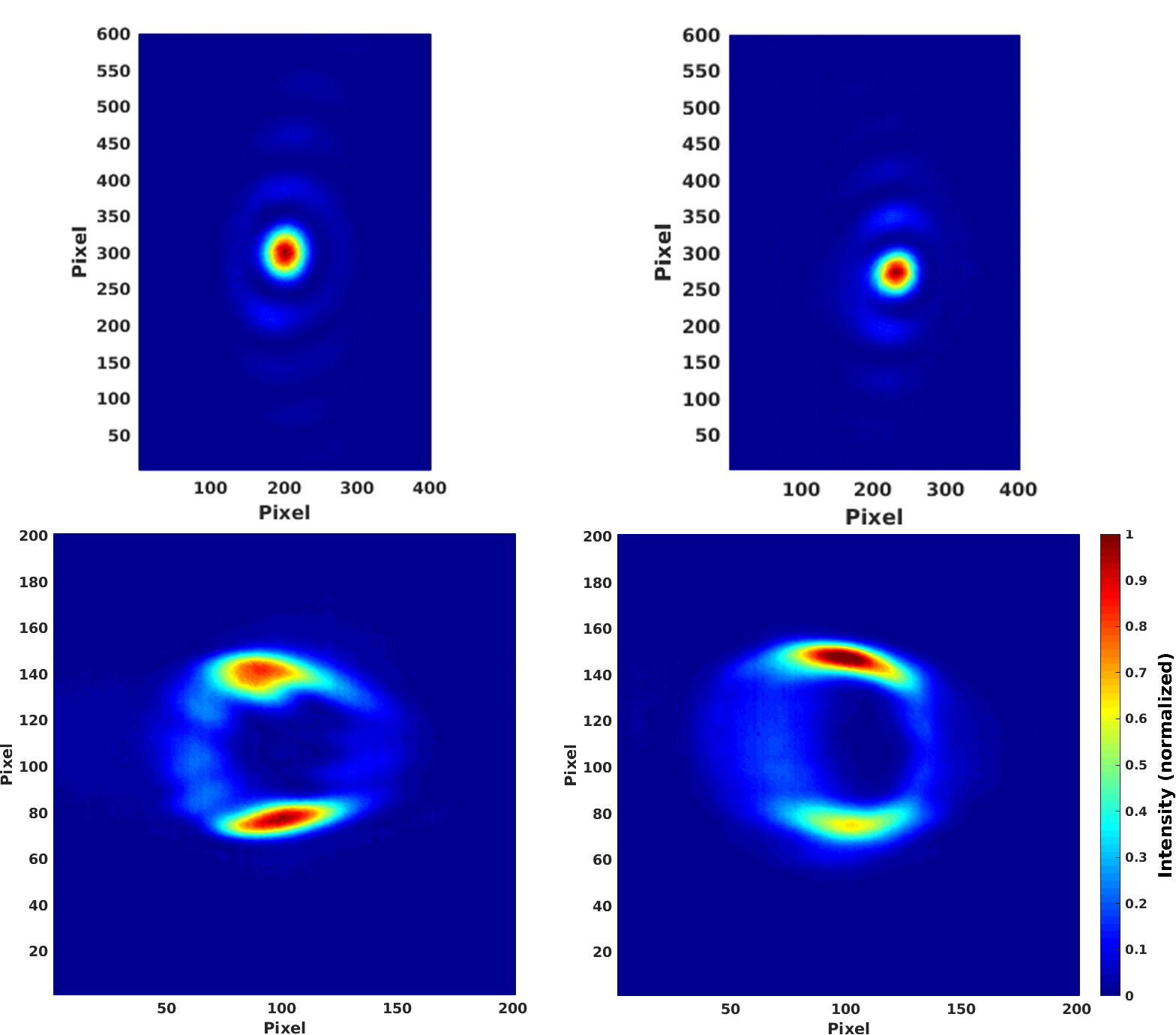}
\caption{Non-Gaussian generated conjugate (top left) and output probe (top right) mode profiles.  Typical spatial Fourier transform patterns of a generated conjugate (bottom left) and output probe (bottom right) mode.}
\end{figure}

Figure 2 shows the spatial profile of the pump beam in the interaction region.  In the following discussion, the aperture that is used to compare mode profiles is shown on the right in Fig.\,2, and has an inner diameter of 400 $\mu$m and an outer diameter of 800 $\mu$m.

A full view of the resulting probe and conjugate modes after the 4WM process is shown in the top of Fig.\,3.  It is immediately evident that the output modes are non-Gaussian, and are constrained to be along the phase-matching cone of the 4WM process.  After collimating the output probe and conjugate modes, we investigate their spatial Fourier transforms by placing a lens in their respective paths, and placing a CCD in the Fourier plane \cite{Vaity}.  Typical resulting mode profiles are shown in the bottom of Fig.\,3. While the probe and conjugate mimic each other as expected, they are not perfect mirror images of one another.  This is likely due to the fact that the conjugate beam and probe beam experience different refractive indices while propagating through the Rb vapor, and thus experience slightly different focusing, resulting in different divergences after the 4WM.  However, these are typical images given the parameters mentioned in previous section, and vary only slightly as, for example, the pump power and frequency drift over time (that is, the top two images in Fig.\,3 are taken simultaneously; the bottom, Fourier transform images, require re-alignment of the CCD camera, as well as knife-edges to aid in removing leftover pump light from whichever mode is being imaged).

In order to verify the aperturing mode conversion effect of the 4WM process, we replace the rubidium cell with an annular aperture, and block the pump beam so that only the probe is present.  After the annular aperture a variable slit is introduced in order to simulate the aforementioned asymmetry in the effective interaction time between the horizontal and vertical regions throughout the 4WM process.  The mode is then collimated and Fourier transformed exactly as in the 4WM case. The resulting mode and its Fourier Transform are both shown in Fig.\,4.  The mode matches very well with the mode-converted probe and generated conjugate in the 4WM process case.

\begin{figure}[h]
\centering\includegraphics[width=13cm]{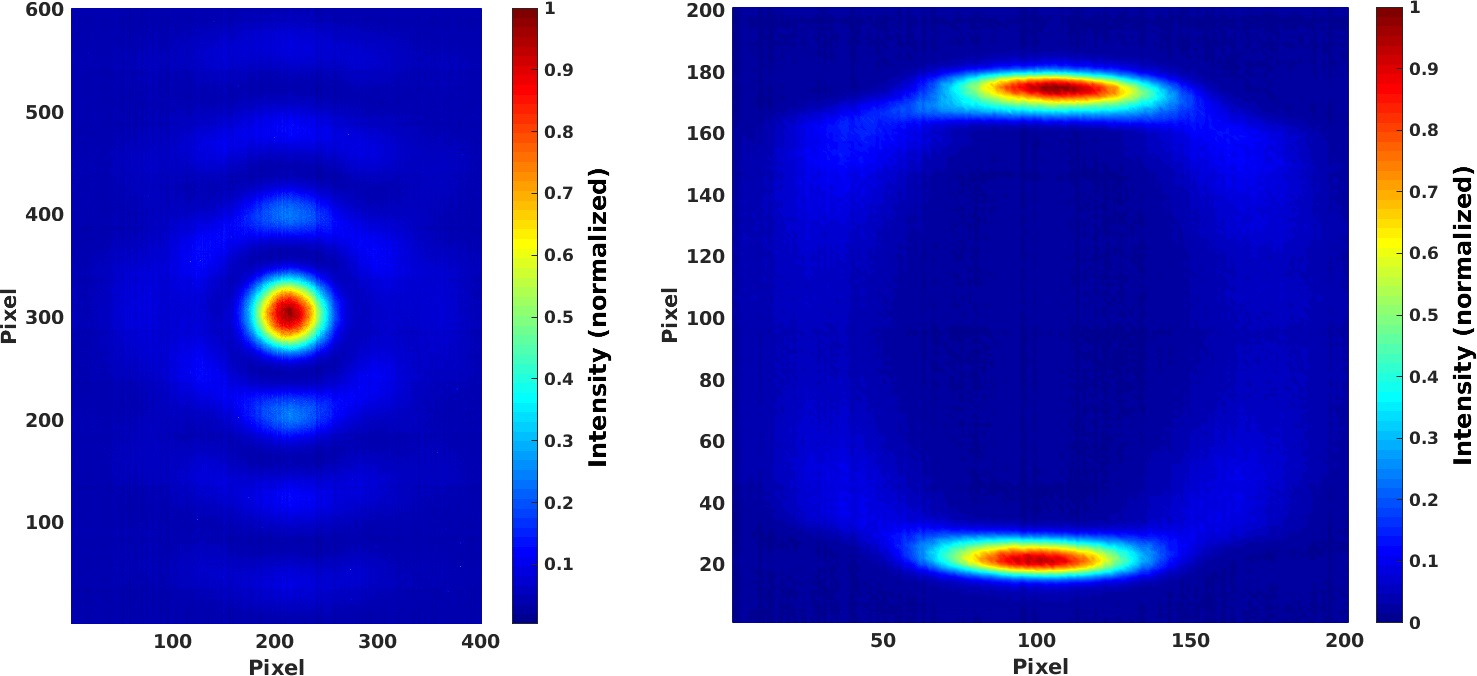}
\caption{Experimentally generated mode when using an annular aperture and variable slit in place of the 4WM interaction (left).  Its Fourier transform (right) agrees well with that of the mode converted probe and generated conjugate as shown in the bottom of Figure 3.}
\end{figure}

We now describe theoretical modeling using standard far-field diffraction theory combined with the asymmetries imposed by finite phase-matching conditions and the non-collinearity of the 4WM process.  Inside the gain medium, pump photons, which form the doughnut shaped beam shown in Fig.\,2, are annihilated in pairs while photons are created in the probe and conjugate modes.  This causes the spatial distribution of the pump mode profile to act as an effective (annular, in the present case) aperture for the probe and conjugate beams.  Given an aperture field distribution $U(\xi,\eta)$, the output diffracted far-field mode profile, under the Fraunhofer approximation, is \cite{introduction_2004}:

\begin{equation}
U(x,y)=\frac{e^{ikz}e^{i\frac{k}{2z}(x^2 + y^2)}}{i\lambda z}\int_{-\infty}^{\infty}\int_{-\infty}^{\infty}U(\xi,\eta)exp\left[-i\frac{2\pi}{\lambda z}(x\xi+ y\eta)\right]d\xi d\eta.
\end{equation}

Here, $\lambda$ is wavelength of the incident light, $k$ is the corresponding wave vector, $z$ is the distance from aperture to the far-field, $x$ and $y$ are the transverse Cartesian coordinate components in the far-field, and $\xi$ and $\eta$ are the Cartesian coordinate components in the plane of the aperture.

An annular aperture of inner radius $a_0$ and outer radius $a_1$ has an amplitude annular transmission function in the aperture-plane coordinates, given in terms of circular functions and a transverse radial coordinate $\rho' = \sqrt{\xi^2 + \eta^2}$, as $t_{A}=g_{annulus}(\rho')= circ(\rho'/a_1) - circ(\rho'/a_0)$. 

For Gaussian laser illumination (field distribution) $U_{Gauss}^{aperture}$ in the aperture plane, the resulting far-field distribution at the lens plane, $U_{lens}$, is given as convolution of Gaussian field distribution and the annular transmission function, which due to the convolution theorem is the multiplication of the Fourier transforms (FT) of each. Once the FT is evaluated at transverse spatial frequency $\nu_{\rho}= \rho/(\lambda z_{lens})$, the field distribution at the lens plane is obtained in terms of the transverse radius $\rho = \sqrt{x^2 +y^2}$, and input beam radius $w(z')$, as: 
\begin{equation}
U_{lens}(\rho)=\frac{e^{ikz_{lens}}e^{i\frac{k}{2z_{lens}}(\rho^2)}}{i\lambda z_{lens}}(U_{Gauss}^{lens}*U_{ring}^{lens})(\rho), \\
\end{equation}
where\\
\begin{equation*}
U_{ring}^{lens}(\rho)=\mathcal{F}\left[g_{annulus}(\rho')\right]=\frac{a_{1}J_{1}(k a_{1}\rho/z) - a_{0}J_{1}(k a_{0}\rho/z)}{\rho/z} 
\end{equation*}
\begin{equation*}
U_{Gauss}^{lens}(\rho)(\rho) = \mathcal{F}\left[U_{Gauss}^{aperture}(\rho')\right]=\mathcal{F}\left[e^{-(\rho'/w(z'))^2}\right]=\frac{w(z')}{\sqrt{2}}e^{-\left(k \rho/\left(2z/kw(z')\right)\right)^2}
\end{equation*}

Thus, the annular diffraction integral contains terms related to first-order Bessel functions of the first kind (Airy-like), $J_1$, as well as that of a Gaussian distribution $U_{ring}$ (since the FT of a Gaussian function is another Gaussian function).

Calculating the spatial Fourier transform (FT) of the resultant modes then allows for direct comparison to the FT'd experimental 4WM and apertured mode cases.  As lenses act as spatial FT devices \cite{introduction_2004} that map the FT of the incoming field at the lens plane, $U_{lens}(\rho)$, to the focal plane, the field distribution at the focal plane is readily obtained from the FT of the incident field with transverse spatial frequencies, $\nu_{\rho_{f}}= \rho_{f}/(\lambda f)$.
\begin{equation}
U_{f}(\rho_{f})=exp(-ikf)exp\left[\frac{ik}{2f}(\rho_{f}^2)\right]\mathcal{F}\left[U_{lens}(\rho)exp\left[\frac{ik}{2f}(\rho^2)\right]\right]_{\nu_{\rho_{f}}}
\end{equation}

Excluding multiplicative phase factors, the diffraction integral is a 2D FT of the field at the lens plane, $\mathcal{F}_{2D}\left[U(\xi,\eta)\right]$, with spatial frequencies, $\nu_{x}=x/\lambda z$, and, $\nu_{y}=y/\lambda z$.

\begin{figure}[t]
\centering\includegraphics[width=8cm]{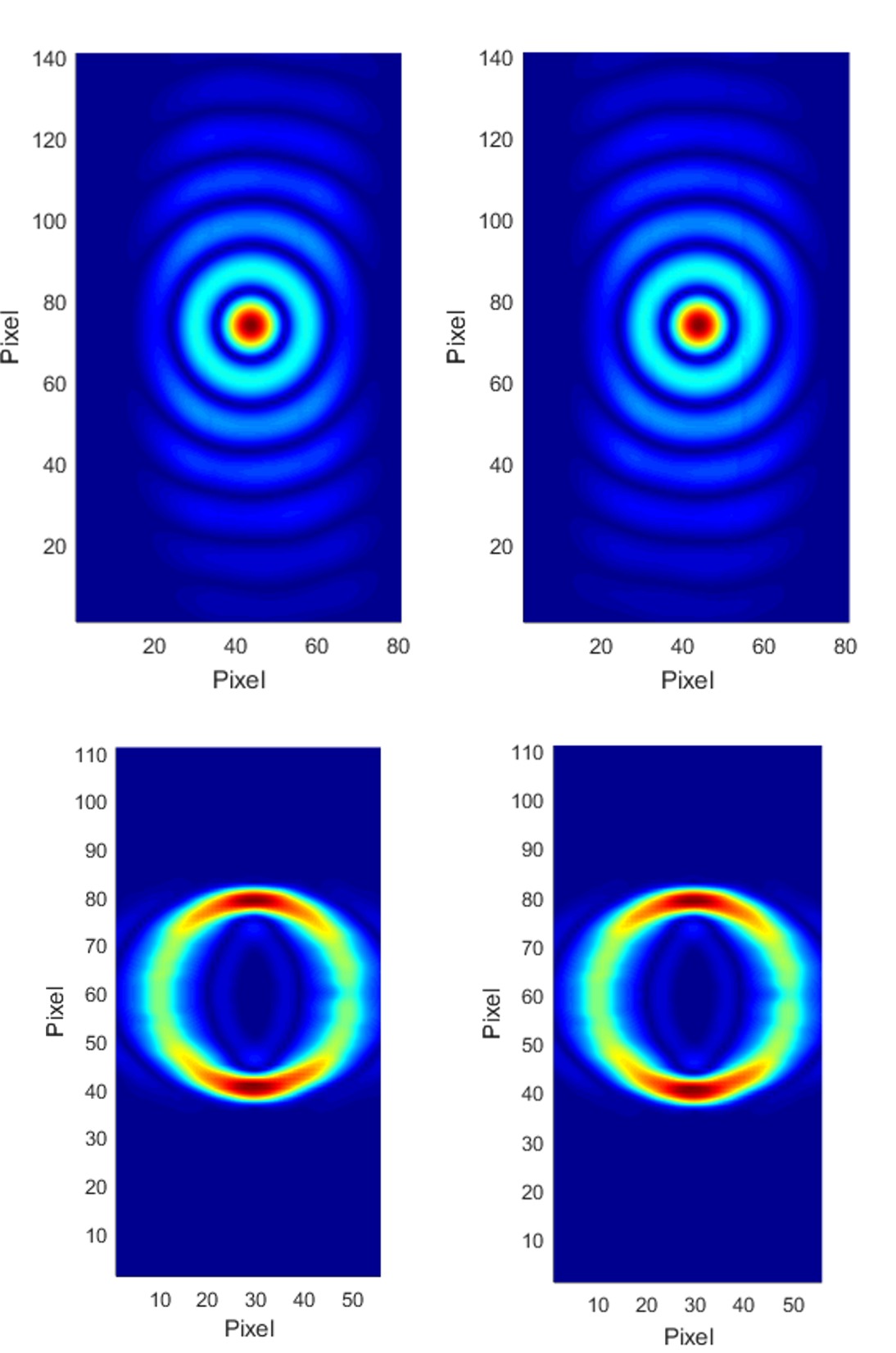}
\caption{Theoretical modeling to simulate the resulting mode profile when a Gaussian mode is incident on an annular aperture, and constrained by the phase-matching conditions of the 4WM process for the conjugate (top left) and probe modes (top right).  Their spatial Fourier transforms are shown below each image.}
\end{figure}

As seen in Fig.\,3, the two 4WM output modes are not perfectly circularly symmetric, as ideal diffraction theory would predict for an annular aperture.  Two important aspects of the 4WM aperturing that contribute to this are the finite phase-matching bandwidth of the processs, as well as the spatial asymmetry in the effective interaction time that the probe and conjugate experience.  When perfect phase-matching is achieved, the probe and conjugate beams experience maximal gain (or, maximal transmission when described as an aperture).  Nevertheless, for a small geometric phase-mismatch, $\Delta k=2k_0-k_{p}-k_{c}$, finite gain is tolerated inside a conical projection (ring) area whose thickness is determined by $\Delta k$.  This can easily be visualized when the nonlinear process is in the extremely high-gain (effectively infinite) regime, where self-seeded (spontaneous) 4WM results in a cone of generated light \cite{narum_1981,Harter_82,erez_1986,Pender_90,valley_1990,zhifan}.  The probe and conjugate beams are effectively bound by this cone, which keeps the Bessel-Gauss-like patterns inside a ring area (two dimensional projection of the cone). 

To account for this asymmetry, we apply a soft spatial-filter to the calculated mode profiles given by Eq.\,2 (or rather, the intensities, which is the modulo squared of Eq.\,2), that corresponds to the the directions with which finite phase-mismatch is tolerated in the 4WM process.  The analytical formulae for the spatial dependence of the 4WM efficiency on the phase-mismatch parameter, $\Delta k$, which is explicitly derived in \cite{alberto2013} for both the probe and conjugate modes is:
\begin{equation}
g_{pr}=\left|exp(\delta a L)\left[\textrm{cosh}(\varepsilon L)+\frac{a}{\varepsilon}\textrm{sinh}(\varepsilon L)\right]\right|^2 
\end{equation}

\begin{equation}
g_{conj}=\left|exp(\delta a L)\frac{a_{cp}}{\varepsilon}\textrm{sinh}(\epsilon L)\right|^2  
\end{equation}
Here $L$ is the interaction length, $\delta a=(a_{pp} - a_{cc} + i\Delta k)/2$, where $\Delta k$ is the geometric phase-mismatch (which provides the spatial dependence of the applied soft aperture here), and $\varepsilon$, $a_{pp}$, and $a_{cc}$ are parameters that depend on the direct and cross-susceptibilities of the medium, as discussed in detail in \cite{alberto2013}. This spatial variation for the probe and conjugate modes is normalized such that peak efficiency corresponds to a transmission of one, and is then multiplied by the intensity of the apertured mode (modulus square of Eq.\,(2)), resulting in the theoretical probe and conjugate intensity distributions shown in Fig.\,5 (top).

We then calculate the effect of Fourier transforming the resultant modes via a lens.  Numerical simulations resulting from Eq.\,(2), and their corresponding FTs (Eq.\,(3)), after being spatially-filtered by a soft aperture corresponding to the finite phase-matching bandwidth of the 4WM process (as described above), are shown in the bottom of Fig.\,5.  These simulations confirm that the 4WM process behaves as an effective aperture, and agree well with the experimental data, both for the 4WM output modes, as well as the apertured modes (Fig.\,4).

\begin{figure}[t]
\centering\includegraphics[width=13cm]{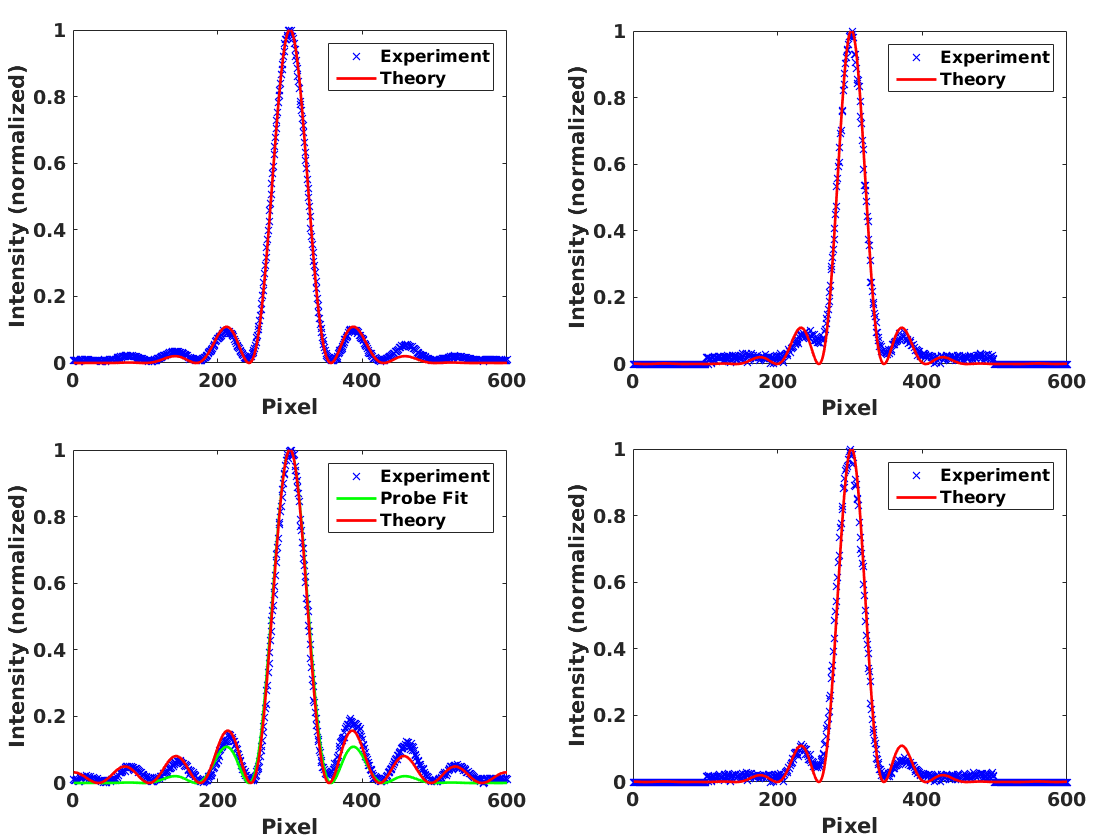}
\caption{Vertical (left) and horizontal (right) slices through the experimentally-generated 4WM output conjugate (top) and probe (bottom) modes shown in the top of Fig.\,3.  The average of ten-pixel wide slices are used in all cases.  Blue x's are experimental data all solid lines are the theoretical fits.  All theoretical fits have a value of $\epsilon = 0.55$, with the exception of the probe vertical slice (bottom left, red line fit has a value of $\epsilon = 0.75$), as discussed in the text.  The green line on the bottom left plot is again the theoretical fit for a value of $\epsilon = 0.55$, shown for comparison.}
\end{figure}

As a final measure of the spatial-mode structure of the 4WM output probe and conjugate modes, we wish to compare their radial intensity distsributions to those resulting from far-field diffraction theory of a simple annular aperture.  While the output 4WM modes are not radially-symmetric due to phase-matching constraints, we find that by taking vertical and horizontal slices through the probe and conjugate modes, the distributions agree qualitatively with the intensity distribution along a given radial direction due to Fraunhofer diffraction through an annular aperture, which is analytically given as \cite{sheppard1977use}:
\begin{equation}
I(v) = \frac{1}{(1-\epsilon ^{2})^{2}}\Big[ \frac{2J_{1}(r)}{v}-\epsilon ^{2}\Big(\frac{2J_{1}(\epsilon r)}{\epsilon v}\Big)\Big]^{2}I_{0},\\
\end{equation} 
where $r$ is the radial direction, $J_{1}$ is a Bessel function of first order, $I_{0}$ is the intensity at the focal point (which we use as a parameter simply to normalize the peak intensities), and $\epsilon$ is a parameter that quantifies the ratio of the inner-to-outer radii of the annular aperture.  For the data shown in Fig.\,6, which contains slices that are 10 pixels wide through the 4WM data shown in the top of Fig.\,3, this value is $\epsilon = 0.55$ for both conjugate plots, the probe horizontal plot, as well as the green fit on the vertical probe plot (bottom left).  This value is close to that of the annular aperture ($\epsilon = 0.5$) used in the experiment, whose resultant modes closely mimicked those of the resultant 4WM non-Gaussian modes.  The red line fit on the probe vertical slice plot (bottom left in Fig.\,3) is for a value of $\epsilon = 0.75$.  We emphasize here that these theoretical fits do not account for phase-matching, nor gain in the 4WM process.  This analysis is to demonstrate that the spatial-mode structure of the output 4WM modes is well-approximated by those generated by sending light through a simple annular aperture, thus reinforcing the idea that the 4WM process is acting as an effective aperture, under the present experimental parameters of a non-Gaussian pump beam that is smaller in spatial extent than the incident Gaussian probe mode.  As we are primarily interested in the spatial structure of the modes, the peak intensities of the experimental data and theoretical distributions are normalized.  We speculate that in addition to phase-matching constraints, one of the primary causes of imperfect agreement with theory in this section is that the 4WM pump beam is not a hard-aperture, whereas Eq.\,4 assumes a perfectly hard-aperture that only allows light to pass through in between the inner and outer radii of an annulus.

\section{Conclusion}
We have shown that the non-degenerate, non-collinear four-wave mixing process in warm atomic vapor may be used as a mode converter for optical beams.  The process amplifies the input mode, rather than attenuate it, and simultaneously generates a second, conjugate mode that exhibits a similar profile to that of the output probe mode.  When the pump beam in such experiments is smaller in spatial extent than the input probe beam, the 4WM process behaves as an effective aperture, resulting in the mode conversion and non-Gaussian mode generation.  Results from the 4WM process are shown to agree well with both the experimental and theoretical output mode profiles that result from a Gaussian mode incident on an annular aperture that is shaped like the pump beam, given the limitations of finite phase-matching and mode overlap throughout the process.  We speculate that this effect is not limited to the 4WM system, but is applicable to any spatially-multimode gain system.

\section*{Acknowledgments} 
This work was made possible by support from Northrop Grumman - \textit{NG NEXT}, as well as the Louisiana State Board of Regents.

\section{References}
\bibliographystyle{refstyle}

\bibliography{references_v1}

\end{document}